\documentclass[12pt]{article}
\usepackage{epsf}
\title{ {\bf Lepton flavor violating Higgs
decays and unparticle physics}}
\author{\vspace{1cm}\\
        {\bf E. O. Iltan,}
        \thanks{E-mail address:
        eiltan@newton.physics.metu.edu.tr}
 \\
        Physics Department, Middle East Technical University \\
        Ankara, Turkey\\}

\date{}

\begin{document}
\setlength{\baselineskip}{24pt}
\maketitle
\setlength{\baselineskip}{7mm}
\begin{abstract}
We predict the branching ratios of the lepton flavor violating
Higgs decays $H^0\rightarrow e^{\pm} \mu^{\pm}$, $H^0\rightarrow
e^{\pm} \tau^{\pm}$ and $H^0\rightarrow \mu^{\pm} \tau^{\pm}$ in
the case that the lepton flavor violation is carried by the scalar
unparticle mediation. We observe that their branching ratios are
strongly sensitive to the unparticle scaling dimension and  they
can reach to the values of the order of $10^{-4}$, for the heavy
lepton flavor case and for the small values of the scaling
dimension.
\end{abstract}
\thispagestyle{empty}
\newpage
\setcounter{page}{1}
%
The hunt of the Higgs boson $H^0$ in Large Hadron Collider (LHC)
is one of the main goals of physicists to test the standard model
(SM), to get strong information about the mechanism of the
electroweak symmetry breaking, the Higgs mass and to determine the
scale of the new physics beyond. From theoretical point of view,
the couplings of Higgs boson with the fundamental particles are
well defined and the branching ratios (BRs) of its various decays
have been estimated as a function of the Higgs boson mass. There
are predictions on the Higgs mass limits from coupling to
$Z/W_\pm$ $m_{H^0}>114.4\, CL \%95 $ and, indirect one, from
electroweak analysis $m_{H^0}=129^{+74}_{-49}$ \cite{PDG2008}
($m_{H^0}=114^{+69}_{-45}$ \cite{LEP2004}).

The present work is devoted to the analysis of the lepton flavor
violating (LFV) Higgs boson decays in an appropriate range,
$110-150\, (GeV)$, of the Higgs boson mass. There are various
analysis done on LFV Higgs boson decays in the literature. In
\cite{Cotti, Marfatia} $H^0\rightarrow \tau\mu$ decay has been
studied in the framework of the 2HDM. In \cite{Cotti}, large $BR$,
of the order of magnitude of $0.1-0.01$, has been obtained and in
\cite{Marfatia}, the $BR$ was obtained in the interval
$0.001-0.01$ for the Higgs mass range $100-160\, (GeV)$, for the
LFV parameter $\lambda_{\mu\tau}=1$. \cite{Koerner} is devoted to
the observable CP violating asymmetries in the lepton flavor (LF)
changing $H^0$ decays with $BR$s of the order of
$10^{-6}-10^{-5}$. The LFV $H^0\rightarrow l_i l_j$ decay has been
studied also in \cite{Assamagan}, in the framework of the two
Higgs doublet model type III. In these works the LF violation is
carried by the lepton-lepton-new Higgs boson couplings which are
free parameters of the model used. In our analysis we consider
that the LF violation is carried by the scalar unparticle
($\textit{U}$)-lepton-lepton vertex and the scalar unparticle
appears in the internal line in the loop.

The unparticle idea, which is based on the interaction of the SM
and the ultraviolet sector, having non-trivial infrared fixed
point at high energy level, is introduced by Georgi \cite{Georgi1,
Georgi2}. Georgi considers that the ultraviolet sector comes out
as new degrees of freedom, called unparticles, being massless and
having non integral scaling dimension $d_u$, around,
$\Lambda_U\sim 1\,TeV$. The interactions of unparticles with the
SM fields in the low energy level is defined by the effective
lagrangian
\begin{equation}
{\cal{L}}_{eff}\sim
\frac{\eta}{\Lambda_U^{d_u+d_{SM}-n}}\,O_{SM}\, O_{U} \,,
\label{efflag}
\end{equation}
where $O_U$ is the unparticle operator, the parameter $\eta$ is
related to the energy scale of ultraviolet sector, the low energy
one and the matching coefficient \cite{Georgi1,Georgi2,Zwicky} and
$n$ is the space-time dimension.

In literature, the unparticle effect in the processes, which are
induced at least in one loop level, is studied in various works
\cite{Lenzz}-\cite{TMAliev}. The process we study exists at least
in one loop level and the effective interaction lagrangian, which
drives the LFV decays in the low energy effective theory, reads
\begin{eqnarray}
{\cal{L}}_1= \frac{1}{\Lambda_U^{du-1}}\Big (\lambda_{ij}^{S}\,
\bar{l}_{i} \,l_{j}+\lambda_{ij}^{P}\,\bar{l}_{i}
\,i\gamma_5\,l_{j}\Big)\, O_{U} \, , \label{lagrangianscalar}
\end{eqnarray}
where $l$ is the lepton field and $\lambda_{ij}^{S}$
($\lambda_{ij}^{P}$) is the scalar (pseudoscalar)  coupling.
Notice that we consider the appropriate operators with the lowest
possible dimension in order to obtain the LFV decays\footnote{The
operators with the lowest possible dimension are chosen since they
have the most powerful effect in the low energy effective theory
(see for example \cite{SChen}).}.

\begin{figure}[htb]
\vskip 3.2truein  \epsfxsize=4.5in
\leavevmode\epsffile{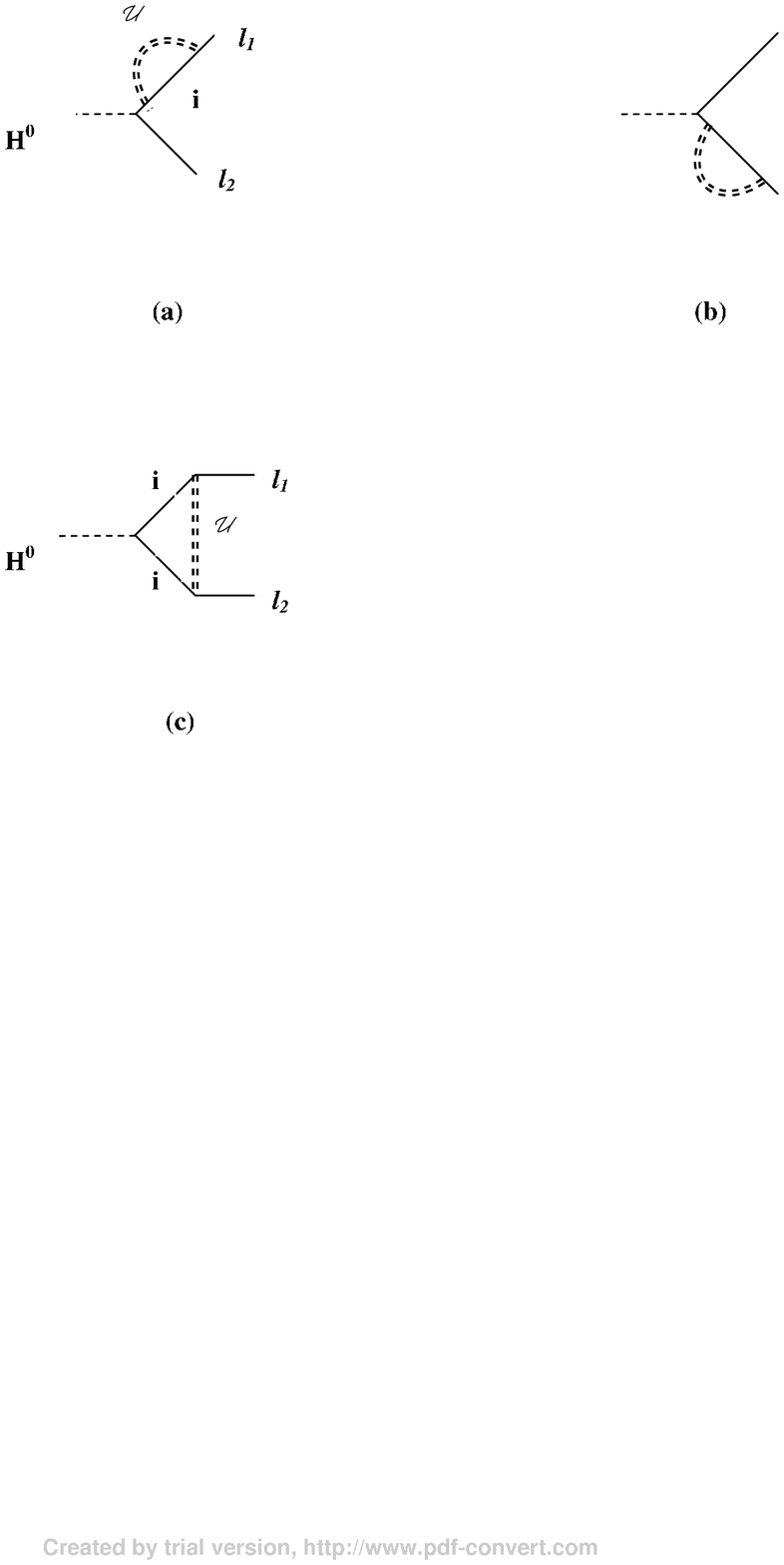} \vskip -4.0truein
\caption[]{One loop diagrams contribute to $H^0\rightarrow
l_1^-\,l_2^+$ decay with scalar unparticle mediator. Solid line
represents the lepton field: $i$ represents the internal lepton,
$l_1^-$ ($l_2^+$) outgoing lepton (anti lepton), dashed line the
Higgs field, double dashed line the unparticle field.}
\label{figselfvert}
\end{figure}

The  $H^0\rightarrow l_1^-\,l_2^+$ decay (see
Fig.\ref{figselfvert}) exists at least in one loop with the help
of the scalar unparticle propagator, which is obtained by using
the scale invariance \cite{Georgi2, Cheung1}:
\begin{eqnarray}
\!\!\! \int\,d^4x\,
e^{ipx}\,<0|T\Big(O_U(x)\,O_U(0)\Big)0>=i\frac{A_{d_u}}{2\,\pi}\,
\int_0^{\infty}\,ds\,\frac{s^{d_u-2}}{p^2-s+i\epsilon}=i\,\frac{A_{d_u}}
{2\,sin\,(d_u\pi)}\,(-p^2-i\epsilon)^{d_u-2} , \label{propagator}
\end{eqnarray}
with the factor $A_{d_u}$
\begin{eqnarray}
A_{d_u}=\frac{16\,\pi^{5/2}}{(2\,\pi)^{2\,d_u}}\,
\frac{\Gamma(d_u+\frac{1}{2})} {\Gamma(d_u-1)\,\Gamma(2\,d_u)} \,
. \label{Adu}
\end{eqnarray}
The function $\frac{1}{(-p^2-i\epsilon)^{2-d_u}}$ in eq.
(\ref{propagator}) becomes
\begin{eqnarray}
\frac{1}{(-p^2-i\epsilon)^{2-d_u}}\rightarrow
\frac{e^{-i\,d_u\,\pi}}{(p^2)^{2-d_u}} \, , \label{strongphase}
\end{eqnarray}
for $p^2>0$ and a non-trivial phase appears as a result of
non-integral scaling dimension.

Now, we present the matrix element square of the LFV $H^0$ decay
(see Fig. \ref{figselfvert} for the possible self energy and
vertex diagrams):
\begin{eqnarray}
|M|^2= 2\Big( m_{H^0}^2
-(m_{l_1^-}+m_{l_2^+})^2\Big)\,|A|^2+2\Big( m_{H^0}^2
-(m_{l_1^-}-m_{l_2^+})^2\Big)\,|A'|^2 \, , \label{Matrx2}
\end{eqnarray}
where
\begin{eqnarray}
A&=&\int^{1}_{0}\,dx\,f_{self}^S+\int^{1}_{0}\,dx\,\int^{1-x}_{0}\,dy\,
f_{vert}^S \, ,\nonumber \\
A'&=&\int^{1}_{0}\,dx\,f_{self}^{\prime\,S}+\int^{1}_{0}\,dx\,
\int^{1-x}_{0}\,dy\, f_{vert}^{\prime\,S} \nonumber \, , \\
\label{funpart}
\end{eqnarray}
and the explicit expressions of $f_{self}^S$,
$f_{self}^{\prime\,S}$, $f_{vert}^S$, $f_{vert}^{\prime\,S}$ read
\newpage
\begin{eqnarray}
f_{self}^S&=& \frac{-i\,c_1\,(1-x)^{1-d_u}}{16\,\pi^2\,\Big(
m_{l_2^+}-m_{l_1^-}\Big)\,(1-d_u)}\,\sum_{i=1}^3\,
 \Big\{(\lambda_{il_1}^S\,
\lambda_{il_2}^S+\lambda_{il_1}^P \lambda_{il_2}^P) \,
m_{l_1^-}\,m_{l_2^+}\,(1-x)\nonumber \\ &\times& \Big(
L_{self}^{d_u-1}-L_{self}^{\prime d_u-1} \Big) -
(\lambda_{il_1}^P\, \lambda_{il_2}^P-\lambda_{il_1}^S
\lambda_{il_2}^S) \,m_i\,\Big(
m_{l_2^+}\,L_{self}^{d_u-1}-m_{l_1^-}\,L_{self}^{\prime d_u-1}
\Big)
\Big\} \, , \nonumber \\ \nonumber \\
f_{self}^{\prime\,S}&=&\frac{i\,c_1\,(1-x)^{1-d_u}}{16\,\pi^2\,\Big(
m_{l_2^+}+m_{l_1^-}\Big)\,(1-d_u)}\,\sum_{i=1}^3\,
 \Big\{(\lambda_{il_1}^P\,
\lambda_{il_2}^S+\lambda_{il_1}^S \lambda_{il_2}^P) \,
m_{l_1^-}\,m_{l_2^+}\,(1-x)\nonumber \\ &\times& \Big(
L_{self}^{d_u-1}-L_{self}^{\prime d_u-1} \Big) -
(\lambda_{il_1}^P\, \lambda_{il_2}^S-\lambda_{il_1}^S
\lambda_{il_2}^P) \,m_i\,\Big(
m_{l_2^+}\,L_{self}^{d_u-1}+m_{l_1^-}\,L_{self}^{\prime d_u-1}
\Big)
\Big\} \, , \nonumber \\ \nonumber \\
f_{vert}^{S}&=& \frac{i\,c_1\,m_i\,(1-x-y)^{1-d_u}}{16\,\pi^2}\,
\sum_{i=1}^3\,\frac{1}{\,L_{vert}^{2-d_u}}\,
 \Bigg\{(\lambda_{il_1}^P\,
\lambda_{il_2}^P-\lambda_{il_1}^S \,\lambda_{il_2}^S)\,
\Big\{(1-x-y)\nonumber
\\&\times&\Bigg( m_{l_1^-}^2\,x+m_{l_2^+}^2\,y
-m_{l_2^+}\,m_{l_1^-}\Bigg)+x\,y\,m_{H^0}^2 -
\frac{2\,L_{vert}}{1-d_u}-m_i^2 \Big\} \nonumber \\&-&
(\lambda_{il_1}^P\,\lambda_{il_2}^P+\lambda_{il_1}^S
\lambda_{il_2}^S)\, m_i\,\Big(
m_{l_1^-}\,(2\,x-1)+m_{l_2^+}\,(2\,y-1)\Big) \Bigg\} \, ,
\nonumber \\ \nonumber \\
f_{vert}^{\prime\,S}&=&\frac{i\,c_1\,m_i\,(1-x-y)^{1-d_u}}{16\,\pi^2}\,
\sum_{i=1}^3\,\frac{1}{\,L_{vert}^{2-d_u}}\,
 \Bigg\{(\lambda_{il_1}^S \,\lambda_{il_2}^P-\lambda_{il_1}^P\,
\lambda_{il_2}^S)\, \Big\{(1-x-y)\nonumber
\\&\times&\Bigg( m_{l_1^-}^2\,x
+m_{l_2^+}^2\,y+m_{l_2^+}\,m_{l_1^-} \Bigg)+x\,y\,m_{H^0}^2 -
\frac{2\,L_{vert}}{1-d_u} \Bigg)-m_i^2 \Big\} \nonumber \\&+&
(\lambda_{il_1}^S
\lambda_{il_2}^P+\lambda_{il_1}^P\,\lambda_{il_2}^S)\, m_i\,\Big(
m_{l_1^-}\,(2\,x-1)+m_{l_2^+}\,(1-2\,y)\Big) \Bigg\} \, ,
\label{spcouplings}
\end{eqnarray}
with
\begin{eqnarray}
L_{self}&=&x\,\Big(m_{l_1^-}^2\,(1-x)-m_i^2\Big)
\, , \nonumber \\
L_{self}^{\prime}&=&x\,\Big(m_{l_2^+}^2\,(1-x)-m_i^2\Big) \, ,
\nonumber \\
L_{vert}&=&(m_{l_1^-}^2\,x+m_{l_2^+}^2\,y)\,(1-x-y)-m_i^2\,(x+y)+m_{H^0}^2\,
x\,y
\, ,
\end{eqnarray}
and
\begin{eqnarray}
c_1&=&\frac{g\,A_{d_u}}{4\,m_W\,sin\,(d_u\pi)\,\Lambda_u^{2\,(d_u-1)}}\,
.
\end{eqnarray}
In eq. (\ref{spcouplings}), the scalar and pseudoscalar couplings
$\lambda_{il_{1(2)}}^{S,P}$ represent the effective interaction
between the internal lepton $i$, ($i=e,\mu,\tau$) and the outgoing
$l_1^-\,(l_2^+)$ lepton (anti lepton). Finally, the BR for
$H^0\rightarrow l_1^-\,l_2^+$ decay can be obtained by using the
matrix element square as
\begin{eqnarray}
BR (H^0\rightarrow l_1^- \,l_2^+)=\frac{1}{16\,\pi\,m_{H^0}}\,
\frac{|M|^2}{\Gamma_{H^0}}\, , \label{BR1}
\end{eqnarray}
with the Higgs total decay width $\Gamma_{H^0}$. In the numerical
analysis,  we consider the BR due to the production of sum of
charged states, namely,
\begin{eqnarray}
BR (H^0\rightarrow l_1^{\pm}\,l_2^{\pm})=
\frac{\Gamma(H^0\rightarrow
(\bar{l}_1\,l_2+\bar{l}_2\,l_1))}{\Gamma_{H^0}} \, .\label{BR2}
\end{eqnarray}
%
%
\\ \\
{\Large \textbf{Discussion}}
\\ \\
In this section, we analyze the BRs of the LFV  $H^0\rightarrow
l_1^- l_2^+$ decays with the assumption that the flavor violation
is induced by the scalar unparticle mediation. The $\textit{U}$-
lepton-lepton vertex drives the LF violation and the decays under
consideration exist in the loop level, in the effective theory. In
the scenario studied there are number of free parameters, namely,
the scaling dimension of the scalar unparticle, the couplings, the
energy scale $\Lambda_u$, the Higgs mass and its total decay
width. These parameters should be restricted by using the current
experimental limits and the mathematical considerations. Here we
choose the scaling dimension in the range $1< d_u <2$, $d_u>1$ not
to face with the non-integrable singularity problem in the decay
rate \cite{Georgi2} and $d_u<2$ to obtain convergent momentum
integrals \cite{Liao1}. For the $\textit{U}$- lepton-lepton
couplings $\lambda^{S(P)}_{ij}$\footnote{We consider that the
scalar $\lambda^{S}_{ij}$ and pseudo scalar $\lambda^{P}_{ij}$
couplings have the same magnitude, namely
$\lambda^{S}_{ij}=\lambda^{P}_{ij}=\lambda_{ij}$.} we consider two
different scenarios:
\begin{itemize}
\item the diagonal couplings $\lambda_{ii}$ respects the lepton
family the hierarchy,
$\lambda_{\tau\tau}>\lambda_{\mu\mu}>\lambda_{ee}$, and the
off-diagonal couplings, $\lambda_{ij}, i\neq j$ are family blind
and universal. Furthermore, we take the off diagonal couplings as,
$\lambda_{ij}=\kappa \lambda_{ee}$ with $\kappa < 1$. In our
numerical calculations, we choose $\kappa=0.5$.
\item
the diagonal $\lambda_{ii}=\lambda_0$ and off diagonal
$\lambda_{ij}=\kappa \lambda_0$ couplings are family blind with
$\kappa=0.5$.
\end{itemize}

The Higgs mass and its total decay width are other parameters
existing in the numerical calculations. The Higgs mass should lie
in a certain range if the SM is an acceptable theory. From the
theoretical point of view, not to face with the unitarity problem
(the instability of the Higgs potential), one considers the upper
(lower) bound as $1.0\,TeV$ $(0.1\, TeV)$ \cite{KHagiwara}. On the
other hand, electroweak measurements results in the prediction of
the Higgs mass  as $m_{H^0}=129^{+74}_{-49}$ \cite{PDG2008}, which
is in the range of theoretical limits. In our numerical
calculation we choose the values $m_{H^0}=110\,(GeV)$,
$m_{H^0}=120\,(GeV)$ and $m_{H^0}=150\,(GeV)$ to observe the Higgs
mass dependence of the BRs of the LFV decays under consideration.
The total Higgs decay width is estimated by using the possible
decays for the considered Higgs mass. The light Higgs boson,
$m_{H^0} \leq 130 \, GeV$, mainly decays into $b \bar{b}$ pair
\cite{Djouadi,MSpira}. However, its detection is difficult due to
the QCD background and the $t\bar{t} H^0$ channel, where the Higgs
boson decays to  $b \bar{b}$, is the most promising one
\cite{Drollinger}. For a heavier Higgs boson $m_{H^0} \sim 180\,
GeV$, the suitable production exist via gluon fusion and the
leading decay mode is $H^0\rightarrow W W \rightarrow l^+ l'^-
\nu_l \nu_{l'}$ \cite{RunII,Dittmar1,Dittmar2}.

Notice that for the energy scale  $\Lambda_u$ we take
$\Lambda_u=10\,(TeV)$ and throughout our calculations we use the
input values given in Table (\ref{input}).
\begin{table}[h]
        \begin{center}
        \begin{tabular}{|l|l|}
        \hline
        \multicolumn{1}{|c|}{Parameter} &
                \multicolumn{1}{|c|}{Value}     \\
        \hline \hline
        $m_e$           & $0.0005$   (GeV)  \\
        $m_{\mu}$                   & $0.106$ (GeV) \\
        $m_{\tau}$                  & $1.780$ (GeV) \\
        $\Gamma (H^0)|_{m_{H^0}=110\,GeV}$   & $0.0026$ (GeV) \\
        $\Gamma (H^0)|_{m_{H^0}=120\,GeV}$   & $0.0029$ (GeV) \\
        $\Gamma (H^0)|_{m_{H^0}=150\,GeV}$   & $0.015$ (GeV) \\
        $G_F$             & $1.16637 10^{-5} (GeV^{-2})$  \\
        \hline
        \end{tabular}
        \end{center}
\caption{The values of the input parameters used in the numerical
          calculations.}
\label{input}
\end{table}

\begin{figure}[htb]
\vskip -3.0truein \centering \epsfxsize=6.8in
\leavevmode\epsffile{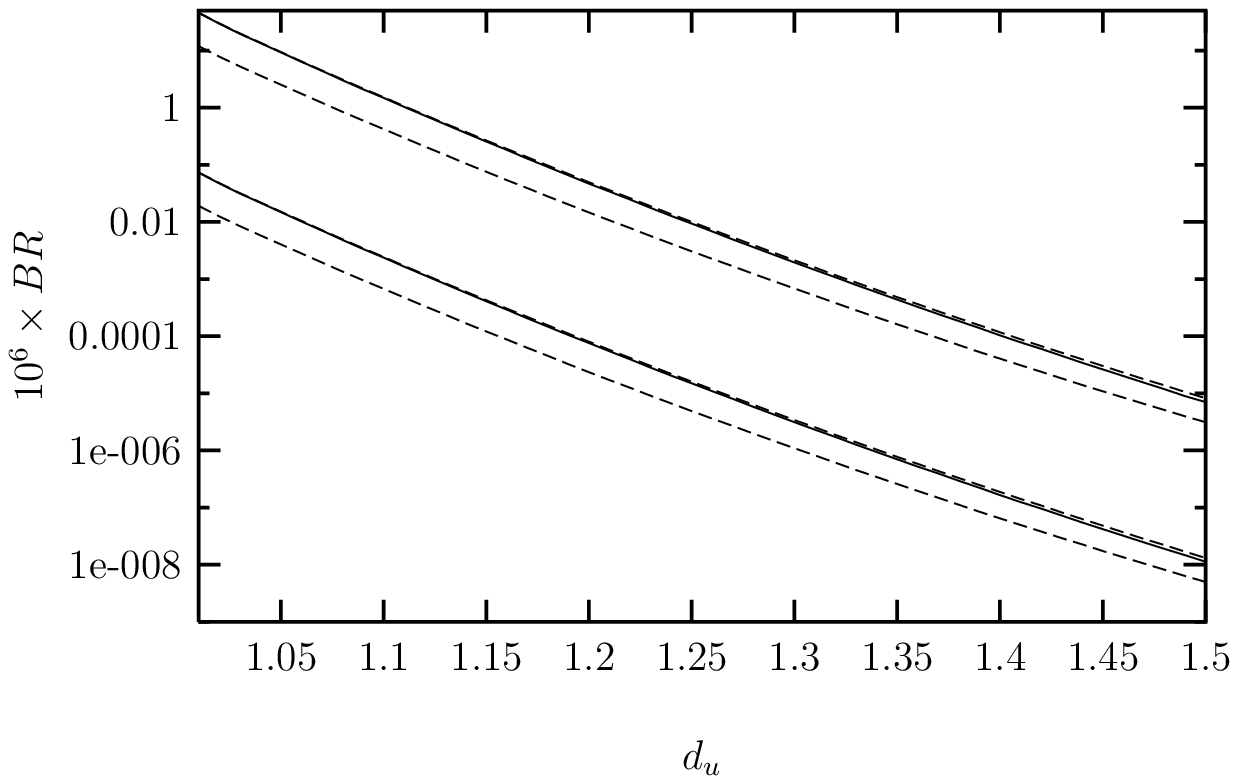} \vskip -3.0truein \caption[]{The
scale parameter $d_u$ dependence of the BR $(H^0\rightarrow
\mu^{\pm}\, e^{\pm})$ for $\Lambda_u=10\, TeV$, the couplings
$\lambda_{ee}=0.1\,\lambda_{\mu\mu}=0.01\,\lambda_{\tau\tau}$.
Here, the lower (upper) solid-dashed line represents the BR  for
$\lambda_{\tau\tau}=10\,(50)$ and $m_{H^0}=110-150\,(GeV)$.}
\label{H0muedu}
\end{figure}

\begin{figure}[htb]
\vskip -3.0truein \centering \epsfxsize=6.8in
\leavevmode\epsffile{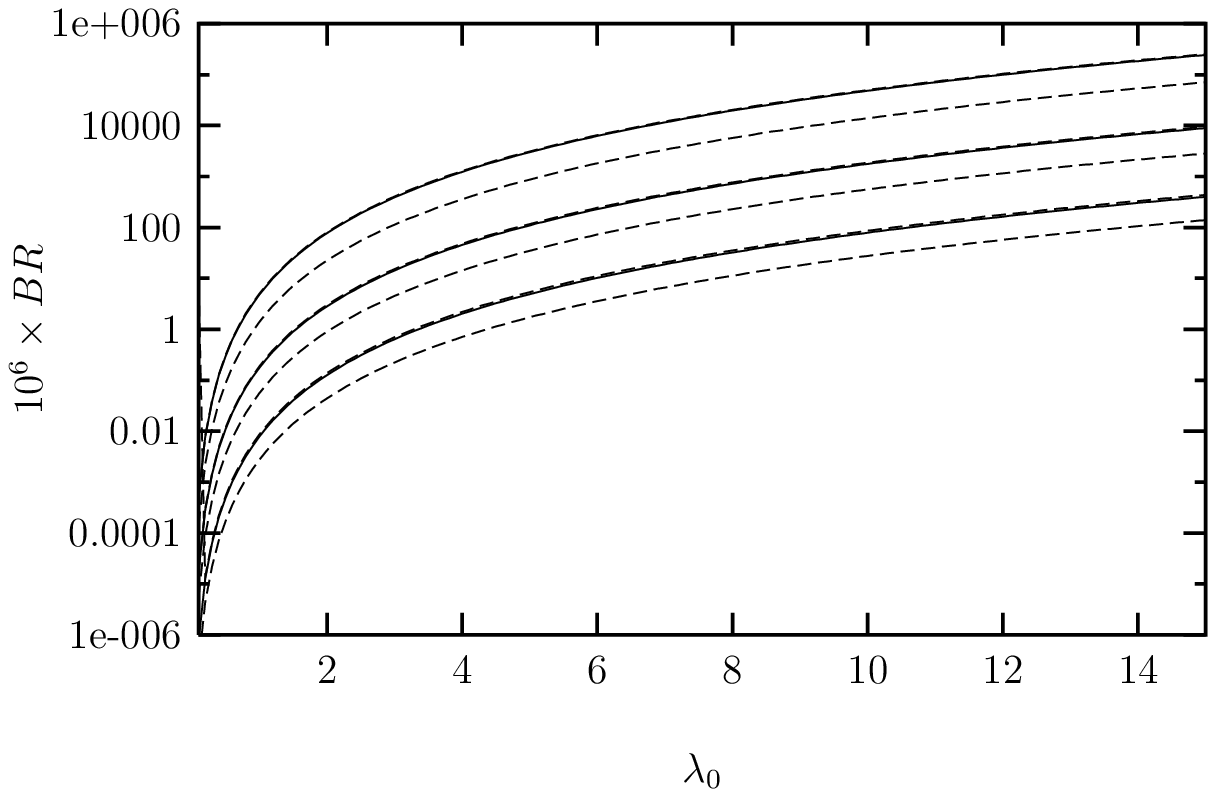} \vskip -3.0truein
\caption[]{$\lambda_0$ dependence of the BR $(H^0\rightarrow
\mu^{\pm}\, e^{\pm})$ for $\Lambda_u=10\, TeV$. Here, the lower
(intermediate, upper) solid-dashed line represents the BR for
$d_u=1.3$ ($d_u=1.2$, $d_u=1.1$) and $m_{H^0}=110-150\,(GeV)$.}
\label{H0muelam}
\end{figure}

In  Fig.\ref{H0muedu}, we present the BR $(H^0\rightarrow
\mu^{\pm}\, e^{\pm})$ with respect to the scale parameter $d_u$,
for the couplings
$\lambda_{ee}=0.1\,\lambda_{\mu\mu}=0.01\,\lambda_{\tau\tau}$.
Here, the lower (upper) solid-dashed line represents the BR  for
$\lambda_{\tau\tau}=10\,(50)$ and $m_{H^0}=110-150\,(GeV)$
\footnote{The BR $(H^0\rightarrow \mu^{\pm}\, e^{\pm})$ for
$m_{H^0}=120\,(GeV)$ is slightly smaller than the one for
$m_{H^0}=110\,(GeV)$ and the difference enhances with the
increasing values of the parameter $d_u$.}. The BR is strongly
sensitive to the scale $d_u$ and, it reaches to the values of the
order of $10^{-6}$ for strong couplings $\lambda_{ij}$, $d_u\sim
1.1$, especially, for the light Higgs boson case.
Fig.\ref{H0muelam} is devoted to the the BR $(H^0\rightarrow
\mu^{\pm}\, e^{\pm})$ with respect to the flavor blind coupling
$\lambda_0$. Here, the lower (intermediate, upper) solid-dashed
line represents the BR for $d_u=1.3$ ($d_u=1.2$, $d_u=1.1$) and
$m_{H^0}=110-150\,(GeV)$. This figure shows that the BR
$(H^0\rightarrow \mu^{\pm}\, e^{\pm})$ enhances considerably even
for the large values of the scale parameter $d_u$, in the case
that the couplings are flavor blind.

\begin{figure}[htb]
\vskip -3.0truein \centering \epsfxsize=6.8in
\leavevmode\epsffile{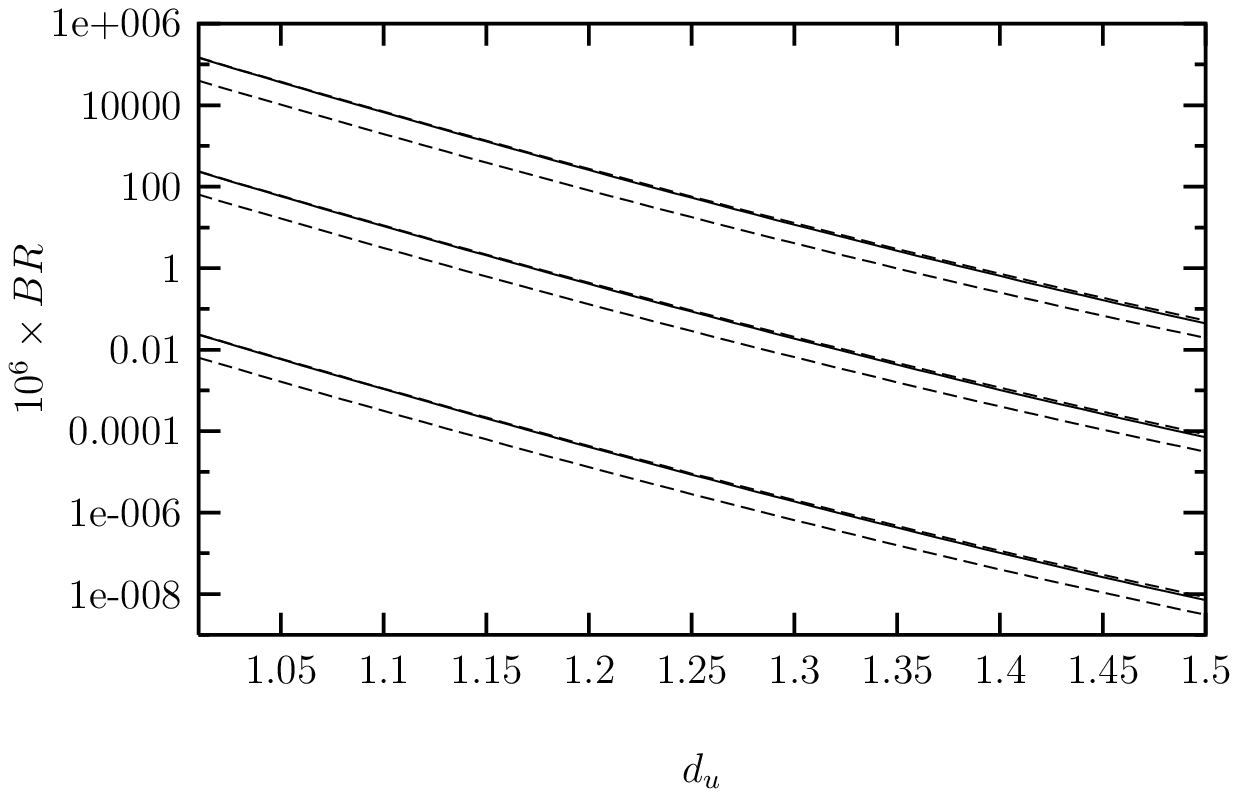} \vskip -3.0truein \caption[]{The
scale parameter $d_u$ dependence of the BR $(H^0\rightarrow
\tau^{\pm}\, e^{\pm})$  for $\Lambda_u=10\, TeV$, the couplings
$\lambda_{ee}=0.1\,\lambda_{\mu\mu}=0.01\,\lambda_{\tau\tau}$.
Here, the lower (intermediate, upper) solid-dashed line represents
the BR for $\lambda_{\tau\tau}=1.0\,(10, 50)$ and
$m_{H^0}=110-150\,(GeV)$.} \label{H0tauedu}
\end{figure}

\begin{figure}[htb]
\vskip -3.0truein \centering \epsfxsize=6.8in
\leavevmode\epsffile{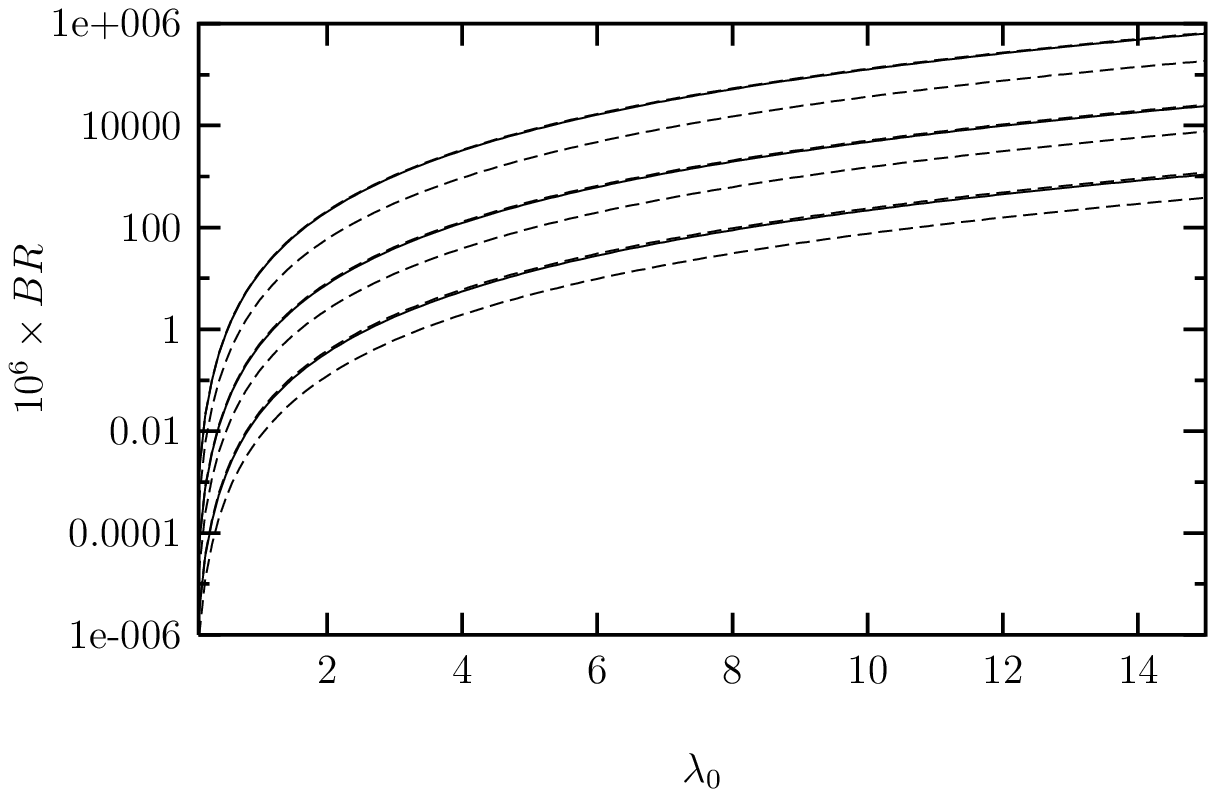} \vskip -3.0truein
\caption[]{$\lambda_0$ dependence of the BR $(H^0\rightarrow
\tau^{\pm}\, e^{\pm})$ for $\Lambda_u=10\, TeV$. Here, the lower
(intermediate, upper) solid-dashed line represents the BR for
$d_u=1.3$ ($d_u=1.2$, $d_u=1.1$) and $m_{H^0}=110-150\,(GeV)$.}
\label{H0tauelam}
\end{figure}

Fig.\ref{H0tauedu}, represents the BR $(H^0\rightarrow
\tau^{\pm}\, e^{\pm})$ with respect to the scale parameter $d_u$,
for the couplings
$\lambda_{ee}=0.1\,\lambda_{\mu\mu}=0.01\,\lambda_{\tau\tau}$.
Here, the lower (intermediate, upper) solid-dashed line represents
the BR for $\lambda_{\tau\tau}=1.0\,(10, 50)$ and
$m_{H^0}=110-150\,(GeV)$. The BR  reaches to the values of the
order of $10^{-6}$ even for weak couplings $\lambda_{ij}$, $d_u
\sim 1.1$. Fig.\ref{H0tauelam} shows the BR $(H^0\rightarrow
\tau^{\pm}\, e^{\pm})$ with respect to the coupling $\lambda_0$.
Here, the lower (intermediate, upper) solid-dashed line represents
the BR for $d_u=1.3$ ($d_u=1.2$, $d_u=1.1$) and
$m_{H^0}=110-150\,(GeV)$. We observe that the BR $(H^0\rightarrow
\tau^{\pm}\, e^{\pm})$ reaches to the values of the order of
$10^{-4}$ even for weak coupling $\lambda_0\sim 1.0 $ in the case
that the couplings are flavor blind.

\begin{figure}[htb]
\vskip -3.0truein \centering \epsfxsize=6.8in
\leavevmode\epsffile{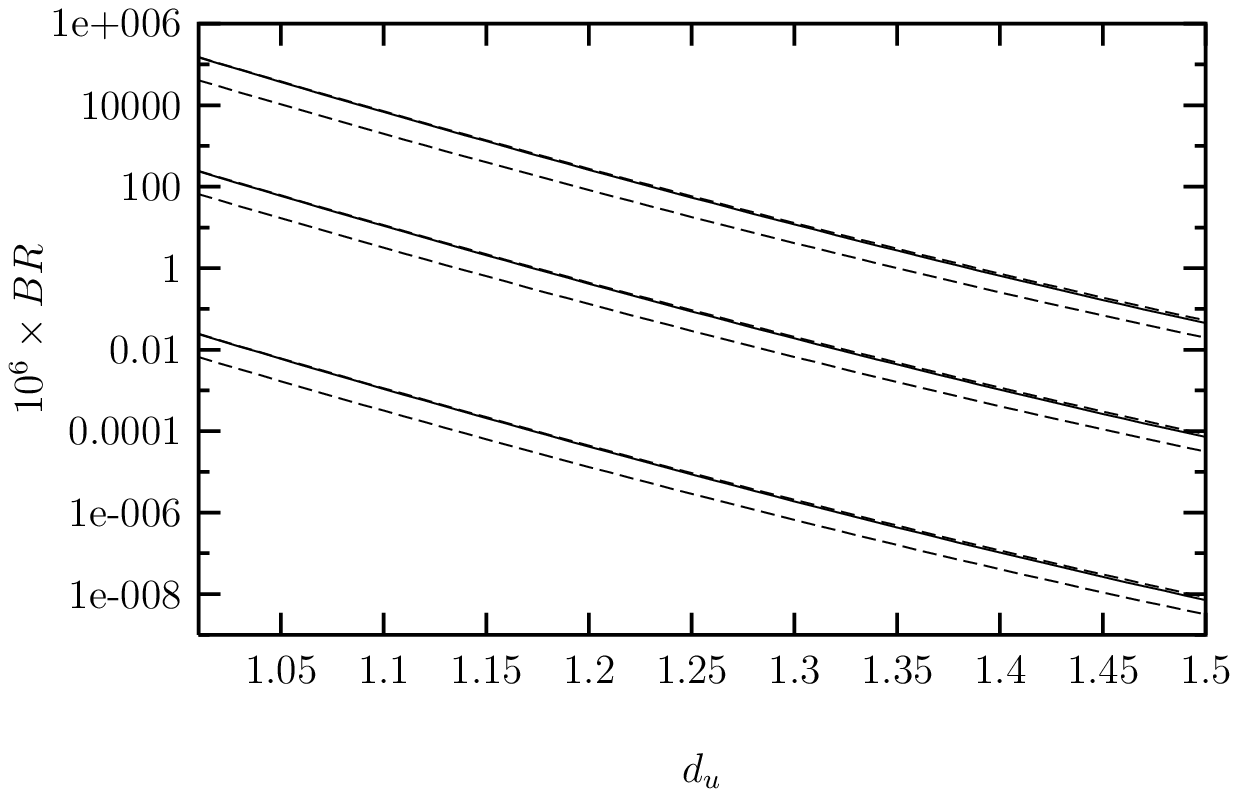} \vskip -3.0truein
\caption[]{The same as Fig.\ref{H0tauedu} but for $H^0\rightarrow
\tau^{\pm}\, \mu^{\pm}$ decay.} \label{H0taumudu}
\end{figure}

\begin{figure}[htb]
\vskip -3.0truein \centering \epsfxsize=6.8in
\leavevmode\epsffile{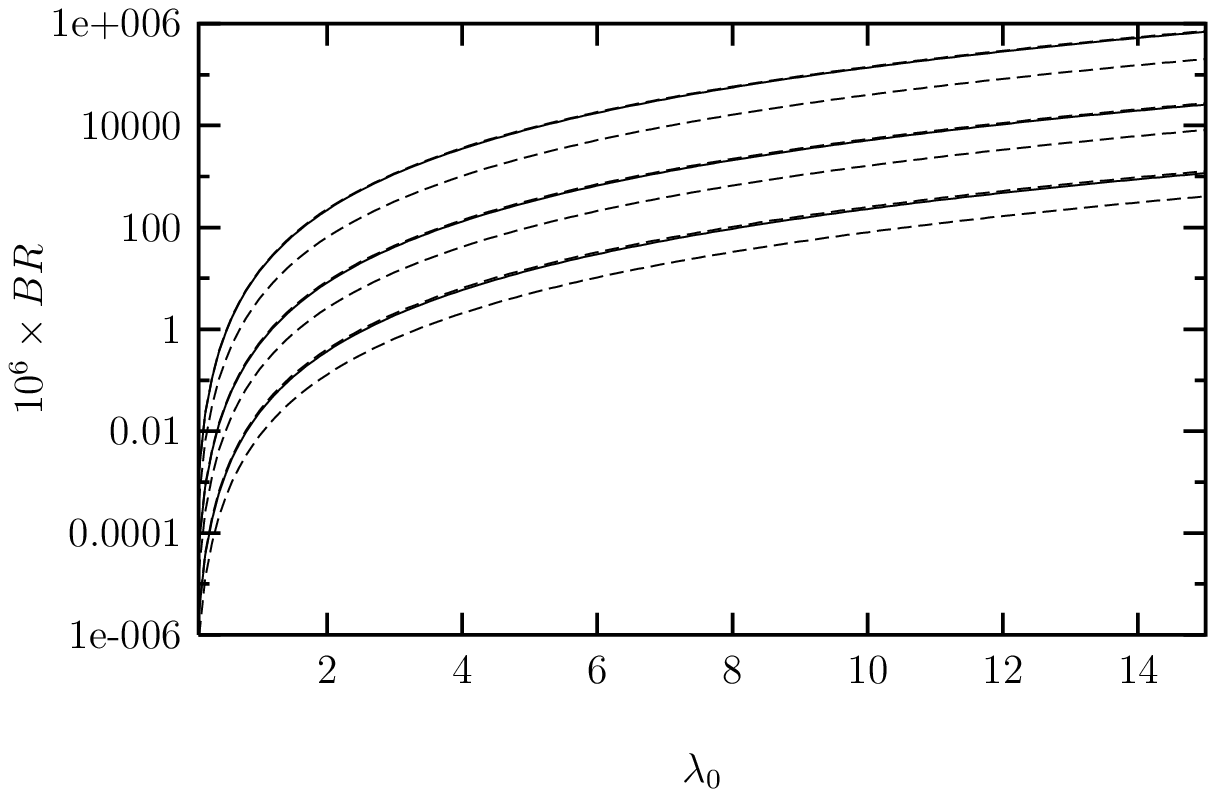} \vskip -3.0truein
\caption[]{The same as Fig.\ref{H0tauelam} but for $H^0\rightarrow
\tau^{\pm}\, \mu^{\pm}$ decay.} \label{H0taumulam}
\end{figure}

Finally, we analyze the the BR $(H^0\rightarrow \tau^{\pm}\,
\mu^{\pm})$ in Figs.\ref{H0taumudu} and \ref{H0taumulam}.
Fig.\ref{H0taumudu} is devoted to the BR $(H^0\rightarrow
\tau^{\pm}\, \mu^{\pm})$ with respect to the scale parameter
$d_u$, for the couplings
$\lambda_{ee}=0.1\,\lambda_{\mu\mu}=0.01\,\lambda_{\tau\tau}$.
Here, the lower (intermediate, upper) solid-dashed line represents
the BR for $\lambda_{\tau\tau}=1.0\,(10, 50)$ and
$m_{H^0}=110-150\,(GeV)$. The BR reaches to the values of the
order of $10^{-6}$ even for weak couplings $\lambda_{ij}$ and $d_u
\sim 1.1$, similar to the $(H^0\rightarrow \tau^{\pm}\,
\mu^{\pm})$ decay. Fig.\ref{H0taumulam} shows the BR
$(H^0\rightarrow \tau^{\pm}\, \mu^{\pm})$ with respect to the
coupling $\lambda_0$. Here, the lower (intermediate, upper)
solid-dashed line represents the BR for $d_u=1.3$ ($d_u=1.2$,
$d_u=1.1$) and $m_{H^0}=110-150\,(GeV)$. It is observed that the
BR $(H^0\rightarrow \tau^{\pm}\, \mu^{\pm})$ can get the values of
the order of $10^{-4}$ for the weak coupling $\lambda_0\sim 1.0$.

As a summary, the LFV decays of the Higgs boson $H^0$ are strongly
sensitive to the unparticle scaling dimension and, for its small
values $d_u < 1.1$, the BRs enhance considerably, especially for
heavy lepton output. In the case that the
$\textit{U}$-lepton-lepton couplings are flavor blind, the BRs of
the decays studied reach to the values of the order of $10^{-4}$
even for weak couplings. The possible production of the Higgs
boson $H^0$ in LHC would stimulate one to study its LFV decays and
the near future experimental results would be instructive to test
the new physics which drives the flavor violation, here is the
unparticle physics.
\newpage
\newpage
\newpage
\end{document}